\documentclass[conference,10pt]{IEEEtran}
\IEEEoverridecommandlockouts
\usepackage{cite}
\usepackage[printonlyused,nolist]{acronym}
\usepackage{amsmath,amssymb,amsfonts,amsthm}
\usepackage{algorithmic}
\usepackage[linesnumbered,ruled,vlined]{algorithm2e}
\usepackage{caption} 
\usepackage{subcaption}
\usepackage{enumitem}
\usepackage{graphicx}
\usepackage{textcomp}
\usepackage{xcolor}
\usepackage[a4paper, total={184mm,239mm}]{geometry}
\def\BibTeX{{\rm B\kern-.05em{\sc i\kern-.025em b}\kern-.08em
    T\kern-.1667em\lower.7ex\hbox{E}\kern-.125emX}}
\usepackage{multirow}
\usepackage{url}
\usepackage[many]{tcolorbox}

\def\BibTeX{{\rm B\kern-.05em{\sc i\kern-.025em b}\kern-.08em
    T\kern-.1667em\lower.7ex\hbox{E}\kern-.125emX}}
\usepackage{pifont}

\begin{document}
\setlength{\intextsep}{1pt}
\setlength{\textfloatsep}{1pt}
\setlength{\abovecaptionskip}{1pt}
\setlength{\belowcaptionskip}{1pt}
\IEEEaftertitletext{\vspace{-1\baselineskip}}

\begin{acronym}
\acro{DNN}[DNN]{Deep Neural Network}
\acro{CNN}[CNN]{Convolutional Neural Network}
\acro{AI}[AI]{Artificial Intelligence}
\acro{LLM}[LLM]{Large Language Model}
\acro{GPU}[GPU]{Graphics Processing Unit}
\acro{SM}[SM]{Streaming Multiprocessor}
\acro{HBM}[HBM]{High-Bandwidth Memory}
\acro{ML}[ML]{Machine Learning}
\acro{MLP}[MLP]{Multi-Layer Perceptron}
\end{acronym}

\title{System-performance and cost modeling of Large Language Model training and inference}


\author{\fontsize{11}{11}\selectfont Wenzhe Guo$^\ast$, Joyjit Kundu$^\ast$, Uras Tos, Weijiang Kong, Giuliano Sisto, Timon Evenblij, and Manu Perumkunnil\\
\fontsize{10}{12}\selectfont Interuniversity Microelectronics Centre (IMEC)\\ Kapeldreef 75,  
3001 Leuven, Belgium}
\maketitle
\def\thefootnote{$\ast$}\footnotetext{These authors contributed equally and are ordered alphabetically}


Large language models (LLMs), based on transformer architectures, have revolutionized numerous domains within artificial intelligence, science, and engineering due to their exceptional scalability and adaptability. However, the exponential growth in LLM size and complexity has outpaced advancements in compute capacity, memory bandwidth, network performance, and cost efficiency, posing significant challenges to their scalability on distributed systems. To address these limitations, alternative model architectures, optimization strategies, communication-aware network topologies, and novel system design approaches have been proposed in literature. This paper introduces a performance-cost modeling methodology for LLM training and inference that integrates state-of-the-art compute techniques with memory optimizations, and latest communication techniques. Building on an analytical performance model, our approach incorporates recent innovations such as the flash attention technique and mixture of experts models to address the memory bandwidth and compute bottlenecks. It also considers the impact of different network topologies and topology-specific communication algorithms with 5D parallellism. The framework also integrates a chiplet cost model. The proposed modeling methodology provides valuable insights to guide future compute system design and facilitates hardware-software co-development, in particular due to its ability to analyze performance-cost trade-offs for various system architectural configurations.

\section{Introduction}
The versatility and scalability of large language models (LLMs) makes them one of the most widely studied artificial intelligence (AI) models in both industry and academia across a wide range of domains (from general machine learning \cite{JMLR:v21:20-074, du-etal-2022-glm, devlin-etal-2019-bert, DBLP:journals/corr/abs-2010-11929, pmlr-v139-touvron21a} to fluid dynamics\cite{mccabe2024multiplephysicspretrainingphysical}, biochemistry\cite{alphafold2021, doi:10.1126/science.ade2574} and materials chemistry \cite{batatia2024foundationmodelatomisticmaterials}). OpenAI's scaling law indicates that model training loss follows a power-law relationship with model size, dataset size, and computational requirements, spanning several orders of magnitude\cite{kaplan2020scalinglawsneurallanguage}. The immense availability of training data has driven a significant increase in model size (\(\sim\) 750x every two years), with model sizes now ranging from billions to trillions of parameters\cite{eff_megatron, DeepSpeedMoE, DeepSpeedInference}, and a corresponding rise in computational demands.

This has also inevitably led to an unprecedented increase in power consumption, costs, and emissions. For instance, training a GPT-3 model is estimated to cost approximately $\$10$ million\cite{wiggers2020}. Thus, performance engineering to optimize existing models \cite{flashattention, flashattention2, du2022glam, eff_megatron} and develop newer and more efficient variants \cite{hymba, DeepSpeedMoE} has become a focus recently. Each technique tries to overcome a limit to the scaling of LLMs:

\textit{1) FlashAttention} alleviates the memory bound nature of attention computation on GPUs by reducing data movement between high-bandwidth memory (HBM) and on-chip SRAM through tiling \cite{flashattention}. Recent developments in this technique have sought to further enhance the performance by leveraging characteristics of the GPU micro-architecture \cite{flashattention2, flashattention3}. 

\textit{2) Mixture-of-Experts} (MoE) models are a new class of models that offer to mitigate prohibitive computation costs of the exponential increase in model and data size \cite{10363447, chen2024energyefficiencylimitstrainingai}. They allow for increased model complexity or size without a corresponding rise in computational costs\cite{GShard}. MoE architecture replaces the multilayer perceptron (MLP) blocks in every alternate transformer layer with sparsely activated expert blocks. As a result, although the parameter count of the MLP block grows with scaling, each token is processed by only a few top experts (typically 1 or 2). This inherent sparsity allows MoE models to achieve high-performance with the cost comparable to a much smaller dense model. However, MoE introduces additional memory overhead for holding multiple experts and communication overhead for routing tokens and results.

\textit{3) Optimizing collective communication on network topologies for multi-dimensional parallelism techniques} has become increasingly important. Multiple parallelization strategies like data parallelism (DP), model tensor parallelism (TP), pipeline parallelism (PP), expert paralleilism (EP), sequencue parallelism (SP) are used to trade-off their (dis)advantages to optimize the distributed training or inference of LLMs at scale. Using the optimal technique for the model, and using the optimal communication algorithm for the optimal network topology are crucial to minimizing communication overhead \cite{ispass-20-ras}, especially since communication bandwidth scales more slowly than the DRAM bandwidth (~1.4x every 2 years) \cite{AIandMemoryWall}.

\textit{4) Chiplets} are used by manufacturers \cite{nvidia_blackwell_2024, amd_mi300_2023, chiplets_automotive_2023}, to address the slowdown of Moore's law by continuing the scaling of compute systems to meet the growing computational demand. While chiplets lower system cost by improving yield, partitioning the design into multiple chiplets introduces die-to-die interconnects that consume both area and power. Additionally, combining chiplets into one system requires advanced packaging techniques, which come with significant costs as well. This trade-off is important to evaluate very early in system design.

It is clear that sustaining the scalability of LLMs requires optimizations across all aspects of the system. Evaluating possible optimizations at scale is complex and resource-intensive. Therefore, a framework that models the impact of various design choices in software, hardware, and interconnect without the need for actual prototyping is highly desirable.

Our contributions in this work are as follows.
\begin{itemize}
    \item We enhance an analytical performance modeling framework for distributed LLMs \cite{arxiv-24-joy} by modeling advanced techniques, namely, FlashAttention, MoE, and topology-specific communication algorithms for parallelism techniques. We validate our new adaptations.
    \item By including expert parallelism, we expand the parallelism space to 5D, enabling for the first time a potential design space exploration of 5D parallelism.
    \item We integrate the performance modeling framework with a chiplet-based cost model \cite{dac-23-gra}. The integration leads to a unified, automated end-to-end framework that enables rapid, large-scale, and consistent exploration of cost-performance trade-offs.
    \item We show use cases combining FlashAttention, MoE, optimal communication algorithms, and cost, highlighting insights in early design decisions for large scale AI systems.
\end{itemize}

\section{Related Work}

In recent years, significant research efforts have been made to study the performance of LLM training and inference. Traditionally, the performance of GPUs and accelerators is studied using cycle-accurate simulation \cite{ispass-09-bak, isca-20-kha} or emulation-based methods \cite{micro-15-ard}. While these approaches are in general difficult to scale for AI workloads \cite{micro-21-ava, vlsi-24-zhu}, domain-specific performance models are proposed to provide insights into the system design trade-offs for AI applications \cite{kdd-15, iclr-16, sbac-18, uatc-21, ispass-23-moo}. In this section, we discuss the related work regarding performance models for AI workloads and fabrication cost models of the corresponding systems.


Since scaling language models requires sophisticated mapping or distribution strategies and optimized hardware utilization, several works focus specifically on LLMs. In \cite{sc-2023-isa}, a fast model is developed to aid the architectural design of LLM training and inference systems. It addresses parallelization strategies with respect to data, pipeline, and tensor parallelisms. Extending the parallelization strategies, AMPeD included a simplistic MoE strategy that only considered the all-to-all communication overhead \cite{ispass-23-moo}. In this framework, the memory subsystems, technology implications and network behaviors were not modeled. DeepFlow combined micro-architecture modeling, device mapping, compute graph transformation and performance prediction into one unified framework \cite{deepflow_2024}. It enabled design space exploration across the full stack under resource and power budgets, but is limited to LSTM workloads. Optimus extended the DeepFlow framework for LLM training and inference by modeling the advanced Megatron parallelization strategies, activation recomputations, collective communications and KV caching \cite{arxiv-24-joy}. However, none of these works addresses the impact of FlashAttention \cite{flashattention, flashattention2} or advanced MoE models. Also, they support limited network topologies in the context of network-architecture modeling. A recent study developed an analytical model that explores a complex design space with 4D parallelisms across different transformer architectures \cite{arxiv-24-sha}. It adopted the FlashAttention technique, but its memory model only considered access counts for HBM. Instead, the performance model in this work uses a detailed hierarchical roofline model \cite{arxiv-24-joy} to address the effect of multilevel caching. Moreover, our parallelization engine considers both MoE and FlashAttention. Therefore, our performance model has an enhanced ability to reflect realistic system behavior. In addition, none of the works discussed above models manufacturing costs for studying cost-performance trade-offs.


Regression and learning-based methods emerge as new alternatives for performance modeling. Linear regression is used to predict the training time for CNN in \cite{sbac-18}. Machine learning (ML)-based approaches have also been studied to predict GPU performance and power consumption of the given kernels \cite{taco-21-bra, hpca-15-wug, ia-19-gue}. Gargi et al. \cite{tmpecs-23-ala} systematically study six of these approaches, addressing design choices regarding their accuracy. However, these approaches are difficult to apply for emerging models and systems where measurement data is limited. The prediction model in \cite{arxiv-24-seo} decomposes large DNN kernels into small working sets, then uses MLP to offer precise prediction for unseen neural networks and GPUs. However, they still suffer a common limitation for ML based approaches: they tend to obfuscate the relationships between different system parameters behind the ML models, limiting the ability to provide insights for system designers.

Chiplet technology has been increasingly adopted in the current large-scale AI systems due to its potential in system scaling. However, splitting a large design into smaller chiplets drives up the manufacturing cost. The flexibility of chiplet-based design leads to various possible combinations between integration strategies and system configurations. Therefore, a cost model becomes necessary to evaluate their impact and find the optimal solution.
Stow et al. introduced a cost model for 2.5D/3D integration, emphasizing yield improvements and non-recurring engineering (NRE) cost reduction through modular die partitioning and IP reuse \cite{cost_analysis_2016}. They then compared active and passive interposers, highlighting cost-performance trade-offs, with passive interposers offering cost-effective scalability \cite{cost_eff_2017}. The Chiplet Actuary model expanded cost analysis by including RE and NRE costs, die-to-die interfaces, and packaging reuse \cite{chiplet_actuary_2022}. Graening et al.provides a comprehensive framework to assess various design trade-offs, introducing a quantitative analysis of factors such as defect density, assembly costs, and interconnect overheads, and establishing the optimal chiplet size range under different scenarios \cite{dac-23-gra}. This last model was chosen to integrate in this work for its hierarchical modeling structure, parameterized flexibility and precise cost analysis.

\section{Background \& Methodology}
\subsection{Overview}
\begin{figure*}[h]
    \centering
    \includegraphics[width=0.9\linewidth]{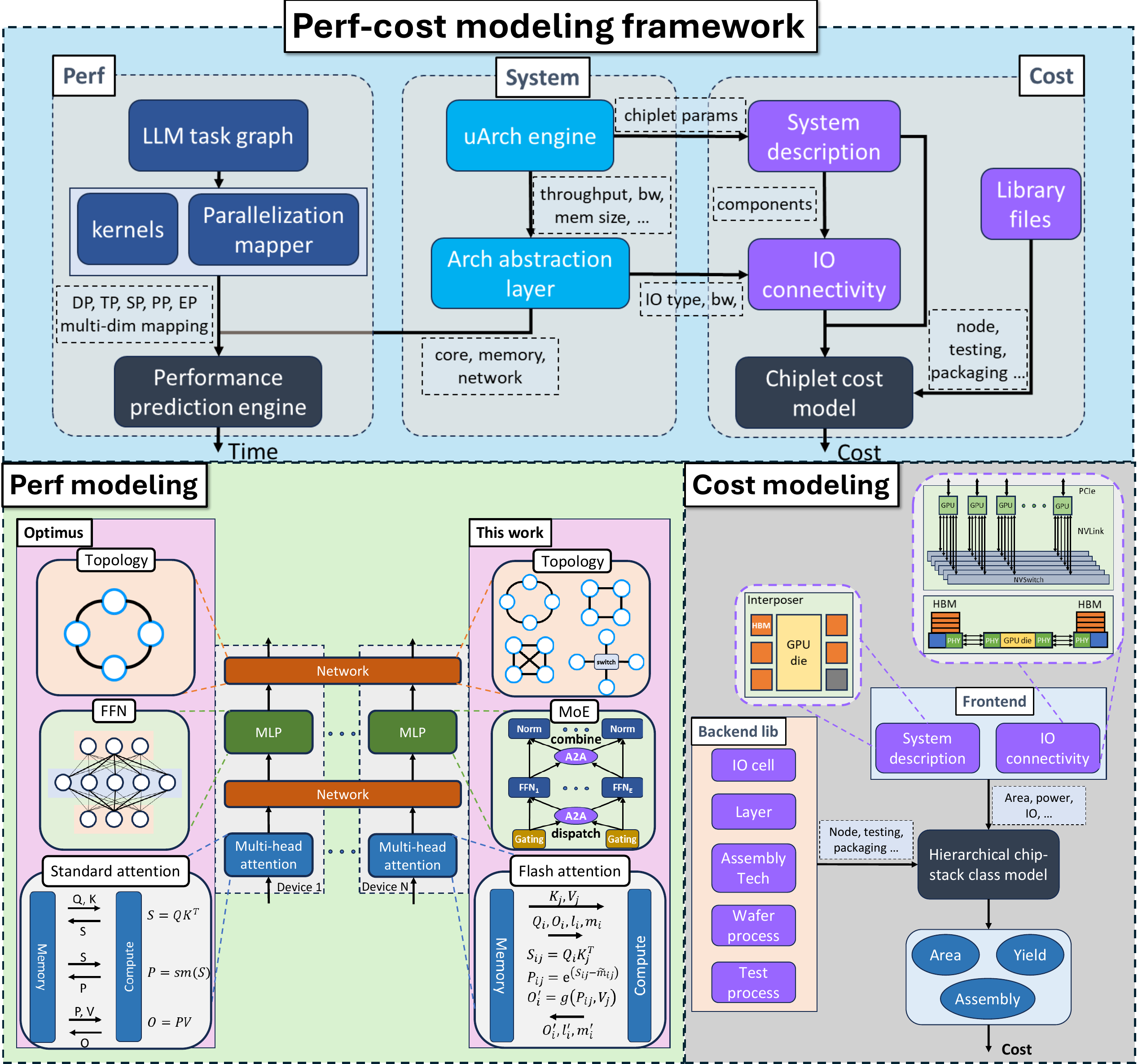}
    \caption{Overview of the proposed framework consisting of a system module, a performance perdiction module and a cost analysis module. The system module uses the microarchitecture ($\mu$Arch) engine and abstraction layer to construct a high-level representation of the system. \textbf{Performance modeling}. Three important components of distributed LLMs are multi-head attention, MLP and networks. Our performance model improves Optimus \cite{arxiv-24-joy} with FlashAttention algorithm, MoE and various network topologies, targeting optimization at each level. \textbf{Cost modeling}. Hierarchical in nature with a nested stack of chiplet objects, frontend configuration files and backend library files. The system description file specifies the organization of chiplets and the IO connectivity file indicates inter-chiplet communication. The total cost considers the area-related cost, assembly cost and yield. }
    \label{fig:framework_overview}
\end{figure*}



We build upon the Optimus \cite{arxiv-24-joy} modeling framework for distributed LLMs, capturing new performance models for state-of-art optimization techniques and constraints relevant to the efficient scaling of LLMs. The Optimus framework includes a collection of the essential implementation techniques for distributed LLMs, including parallelism strategies, activation recomputation, collective communication, grouped query attention, and KV caching. It models a four-dimensional parallelism space, spanning TP, DP, SP and PP. Various PP schedules are modeled in Optimus, including GPipe, PipeDream-Flush and Interleaved-1F1B \cite{eff_megatron}. However, some of the latest advancements are missing. In this work, we incorporate FlashAttention, MoE and topology-aware optimizations for collective communication. Our performance model supports both training and inference. Additionally, we integrate an open-sourced chiplet-based cost model proposed in \cite{dac-23-gra} along with the performance prediction model to enable workload-specific system design trade-offs. We design a proper interface between the system module and the cost model to handle the generation of the required system and IO information, achieving an end-to-end framework. The integration leads to a unified, automated framework that seamlessly blends performance and cost analysis. It enables rapid, large-scale, and consistent exploration of cost-performance trade-offs. These advancements are illustrated in Fig. \ref{fig:framework_overview}.


Our framework consists of three major components, namely, a system module, a performance prediction module (both extended from \cite{arxiv-24-joy}) and a new cost analysis module. In the system module, the micro-architecture engine receives the architecture template of accelerators, essential technology parameters, and resource allocation budgets (e.g., area and power). It then generates the micro-architectural parameters of the accelerator, such as core throughput, memory bandwidth at each level, and network bandwidth. An architecture abstraction layer constructs the representation of concrete accelerators. For known accelerators, parameters can be directly inserted to drive the performance prediction without using the micro-architecture engine (based on DeepFlow\cite{deepflow_2024}).

The performance module builds task graphs for LLM training or inference involving the compute kernels and their dependencies. The task graph is mapped to a distributed system based on certain parallelization strategies. The prediction engine then estimates the computation time via a hierarchical roofline model \cite{deepflow_2024} and the communication time based on collective communication network model. A proper interface is designed between the system module and the cost model to generate the required input information. The cost model can directly extract hardware details of the accelerators and IOs. On combining this with the parameters fetched from the technology library, e.g., cost per unit area, defect density, and assembly cost, the final system-cost can be calculated.

We differentiate our contributions from the base performance modeling framework as shown in the performance modeling block of Fig.~\ref{fig:framework_overview}. As presented, we primarily enable the modeling of FlashAttention, include experts parallelism with MoE models (this expands the design space into five dimensions of parallelism; spanning data parallelism, tensor parallelism, sequence parallelism, pipeline parallelism and expert parallelism), and add a library of network topologies with the corresponding bandwidth optimal algorithms. In addition to this, we integrate a cost model as presented in Fig.~\ref{fig:framework_overview}.

\subsection{Modeling FlashAttention}

The attention mechanism in transformers is primarily memory bound, since the ratio between the flop count and transferred bytes is mostly $\sim \mathcal{O}(1)$. Attention computation involves multiple kernels: {\em skinny} matrix-multiplication (SMM), masking, dropout, and softmax, where the latter three are point-wise operations. Attention incurs $\mathcal{O}(N^2)$ memory accesses, where $N$ is the context length. Thus, it is challenging to scale to larger context lengths. FlashAttention implementation drastically reduces the number of global memory accesses by tiling the data efficiently onto the much faster on-chip SRAM and thus, mitigates this bottleneck\cite{flashattention}. 
The key idea is similar to that of {\em Online softmax} that turns a naive three-pass softmax computation\cite{onlinesoftmax}.


The standard algorithm involves three passes: I. Loading Q, K by blocks from HBM to compute S = Q ${\rm K^T}$ and writing S to back HBM, II. Reading S from HBM to compute P = softmax(S) and writing P to HBM, III. Loading P and V by blocks from HBM to compute O = PV and finally write O back to HBM. 
 Typically, this is a multiple-pass algorithm where each pass above runs for, say, T steps. The second and third passes here require the final result of the preceding pass after T steps, making it nontrivial to fuse all the parts into a single-pass algorithm. The trick is to create surrogates for original sequences where fusing is non-trivial such that the dependency on the T-th term can be bypassed\cite{ZihaoNotes}. 
 This allows to fuse multiple passes into a single-pass kernel involving the introduction of a few surrogate sequences. Here, we describe the high-level equation expressing the execution time for the forward pass closely following the algorithm presented in the first paper\cite{flashattention}. A similar approach is followed for the backward pass. The execution time ($t_{\rm FA}$) is modeled as the following:
 \begin{equation}
      t_{\rm FA} = t_{\rm HBM-ld} + t_{\rm GEMM} + t_{\rm pt-reduc-ops} + t_{\rm HBM-st},
 \end{equation}
where $t_{\rm HBM-ld}$ captures the time to load the Q,K,V, and O matrices along with the surrogate sequences from HBM to SRAM. All the matrices are divided into tiles of size $\sim B \times d$ , where $d$ is the key/value/query length per head and $B$ is determined by the on-chip SRAM size $M$ such that $4B\times d \sim M$ (to fit 4 matrices). $t_{\rm GEMM}$ corresponds to the time spent in the matrix multiplications: S=Q K$^{\rm T}$ and O= P V. 
For a given matrix-multiplication, we calculate the flop count and movement from SRAM to the registers to determine if it is compute throughput bound or SRAM bandwidth bound. We then estimate the execution time using a roofline model. The time spent in element-wise operations like masking, dropout are grouped together with softmax (reduction operation) is expressed as $t_{\rm pt-reduc-ops}$. Finally, the time spent in writing O along with surrogate sequences back to HBM are expressed as $t_{\rm HBM-st}$. The FlashAttention technique drastically reduces $t_{\rm GEMM}$ and $t_{\rm pt-reduc-ops}$ by turning these operations from being memory-bandwidth bound in the standard implementation into compute bound ones. 
 The remaining steps in the task graph along with the MLP block remain unchanged. 

\subsection{Modeling MoE}
Mixture of Experts (MoE) scales transformers by replacing some MLP layers with specialized experts, activating only a subset per token. This increases complexity without extra compute\cite{GShard} but introduces load imbalance and high memory and communication costs. Tokens are routed via softmax-based gating, requiring all-to-all GPU communication to preserve sequence context\cite{arxiv-24-joy}.

We utilize the hybrid Tensor-Expert-Data Parallelism approach in DeepSpeed-TED\cite{Singh_2023}. In this case, the parameterized routing function routes each input token to a unique expert. Thus, the effective computation cost is insensitive to the number of experts and remains the same as that of the base model. In this case the MHA and Experts blocks are parallelized in different fashion. The parallelization identity proposed by DeepSpeed is modeled as follows:
\begin{equation}
    {\rm TP} \times {\rm EP} \times  {\rm DP_{EXP}}
=  {\rm TP} \times  {\rm DP_{MHA}},
\end{equation}
where TP, EP, and DP stand for tensor parallel, expert parallel and data parallel degree respectively. The suffix to DP implies DP degree in the corresponding block (expert or attention). For simplicity we set, ${\rm DP_{EXP}}=1$ (no data parallelism in the expert block), implying  EP = ${\rm DP_{MHA}}$. Hence, the total number of devices is given by, TP $\times$ DP$_{\rm MHA}$. The total number of experts, denoted by $n_{\rm EXP}$, are distributed over EP devices. The computation for each expert is further divided into TP partitions (for details refer to Fig.03 in~\cite{DeepSpeedMoE}). For each expert, the GEMM operation (input I $\times$ weight W$_{\rm FFN1}$ = O$_{\rm FFN1}$) is followed by a non-linear function like GELU. To parallelize efficiently with minimal reduction overhead, W$_{\rm FFN1}$  is column-partitioned, allowing independent application of the non-linearity without synchronization. The resulting output O$_{\rm FFN1}$ is also column-partitioned, requiring the weight matrix W$_{\rm FFN2}$ in the next FFN layer within that expert to be row-partitioned correspondingly. This mapping enables local multiplications, with partial results reduced across devices via a single all-reduce operation in the forward pass, effectively parallelizing both GEMMs within a given expert. For the parallelization details of the MHA block, we refer to \cite{arxiv-24-joy}. 


Following the GShard implementation \cite{GShard}, we modeled the All-to-All communication in the following way, assuming that each device has one expert,
\begin{equation}
    T_{A2A}=\frac{K}{BW_{A2A}}=\frac{ECD\times precision}{BW_{A2A}}
\end{equation}
where $K$ is the data volume to be sent from one device to another, $E$ is the number of experts, $C$ is expert capacity, $D$ is the embedding dimension, and $BW_{A2A}$ is the All-to-All bandwidth of the network. Based on GShard, the use of expert capacity ensures that each expert has its own token buffer of size $C$. The tokens that exceed the buffer are dropped. If the buffer is not full, it will be filled with zeros. So the data volume to be sent is then fixed as $ECM\times precision$, which is not affected by load imbalance. In our experiments, $BW_{A2A}$ is taken from the NVIDIA’s measurements \cite{nvidia2024nccl}.

In addition to the All-to-All communications, we model the time spent in the matrix multiplication operation ($\sim$[N$\times$ D]. [D $\times$ E], where D is the embedding dimension and E is the number of experts) within the local gating function on every device. This is followed by a softmax operation to determine the top expert for every token in the sequence. The GEMM and softmax are modeled using the hierarchical roofline analysis. The rest of the compute or communication remain the same as the base model.  

\subsection{Network Modeling}


This work adopts network modeling functionality that allows the selection of a variety of network topologies. Modeled topologies include ring, switch, fully connected, and 2D mesh network topologies. We also provide the capability to represent scale-out architectures with multi-level hierarchies of the aforementioned topologies. In this manner, it is possible to describe large-scale systems spanning many nodes, with multiple levels of network topologies and networking speeds. 

For all implemented topologies, the total communication delay $D_{\rm total}$ is modeled into three parts, namely serialization delay, link delay, and switching delay. The serialization delay describes the amount of time needed to serialize $d$ bytes over a bandwidth of $b$. Link delay represents the latency incurred by the link itself, i.e. the amount of time it takes for the first bit to arrive at the other end of a network link. This is determined by the number of hops $h$, and the link latency $l$. The switching delay component is optional and depends on the topology in use. If there is more than one hop between a communication source and its destination, this additional switching delay, can be modeled as a constant $D_r$ if needed. Eq. \ref{eq:comm_delay} depicts the total communication delay for point-to-point communication. 

\begin{equation}
D_{\rm total} = \frac{d}{b}+(l \times h)+D_r 
\label{eq:comm_delay}
\end{equation}

Next to the network topology, the collective communication algorithm choice has an equally high impact on the communication delay. In this work, we choose to model a topology-specific collective all-reduction operation for each network topology considered, ensuring a fair comparison between network topologies. Our framework implements the ring reduction algorithm for a ring topology, Rabenseifner's algorithm \cite{doi:10.1177/1094342005051521} for a switch topology, direct reduction for fully connected topology, and a 2D reduction algorithm \cite{kumar2020highlyavailabledataparallel} for a 2D mesh topology. 


\subsection{Cost Modeling}


As chiplet systems are becoming increasingly prevalent in high-performance computing and automotive applications, effective cost modeling is essential due to the wide array of possible combinations between integration techniques and system configurations. This enables cost-performance co-optimization, helping to identify the most optimal configurations for achieving the best balance of performance and cost. We adopt and integrate an existing open-sourced chiplet cost model proposed in \cite{dac-23-gra}, consisting of a hierarchical model with a nested stack of chiplet objects to enable analysis of flexible designs. An overview of the model is illustrated in Fig. \ref{fig:framework_overview}. It relies on front-end system configuration files and a back-end library to perform calculation. The front-end includes a system description and a netlist file describing inter-chiplet connectivity. 
For example, a system resembling a NVIDIA A100 GPU has one reticle-sized GPU chiplet, five HBM2e stacks (each containing 8 DRAM dies and 1 logic die), and one dummy chiplet sitting on the same interposer. It uses a third-generation NVLink (serial) interconnect for GPU-to-GPU connections, and a parallel interconnect for its HBM stacks. The backend library files characterize the various IO types, process nodes, substrates, assembly processes, wafer processes, and test processes, providing specific technological and cost-related details independent from a particular system. 
The model calculates the total cost based on die and assembly cost, expressed as,
\begin{align}
    C =\frac{1}{Y_{\rm assembly}}\left[(\sum_{i=1}^{N_{c}}\frac{C_{\rm die}^i}{Y_{\rm die}^i})+C_{\rm assembly}\right]
\end{align}
where $C_{\rm die}^i$ is the cost of the individual chiplet $i$, $C_{\rm assembly}$ is the corresponding assembly cost, $N_c$ is the number of dies, $Y_{\rm die}^i$ is the yield, and $Y_{\rm assembly}$ is the assembly yield. 
The total area of one chiplet is determined as the greater between the core area plus IO cell area and the area of signal and power pads. 
The assembly cost is modeled as the sum of material cost (e.g., interposer cost) and machine operating cost, which can be extracted from the backend library. 
The negative binomial yield model is applied for the yield calculation \cite{775417}. The cost model has been widely used and validated in various studies \cite{24-ale, 10546772, ECO-CHIP, DRAM-Cost, REED}.
\begin{figure}
    \centering
    \includegraphics[width=\linewidth]{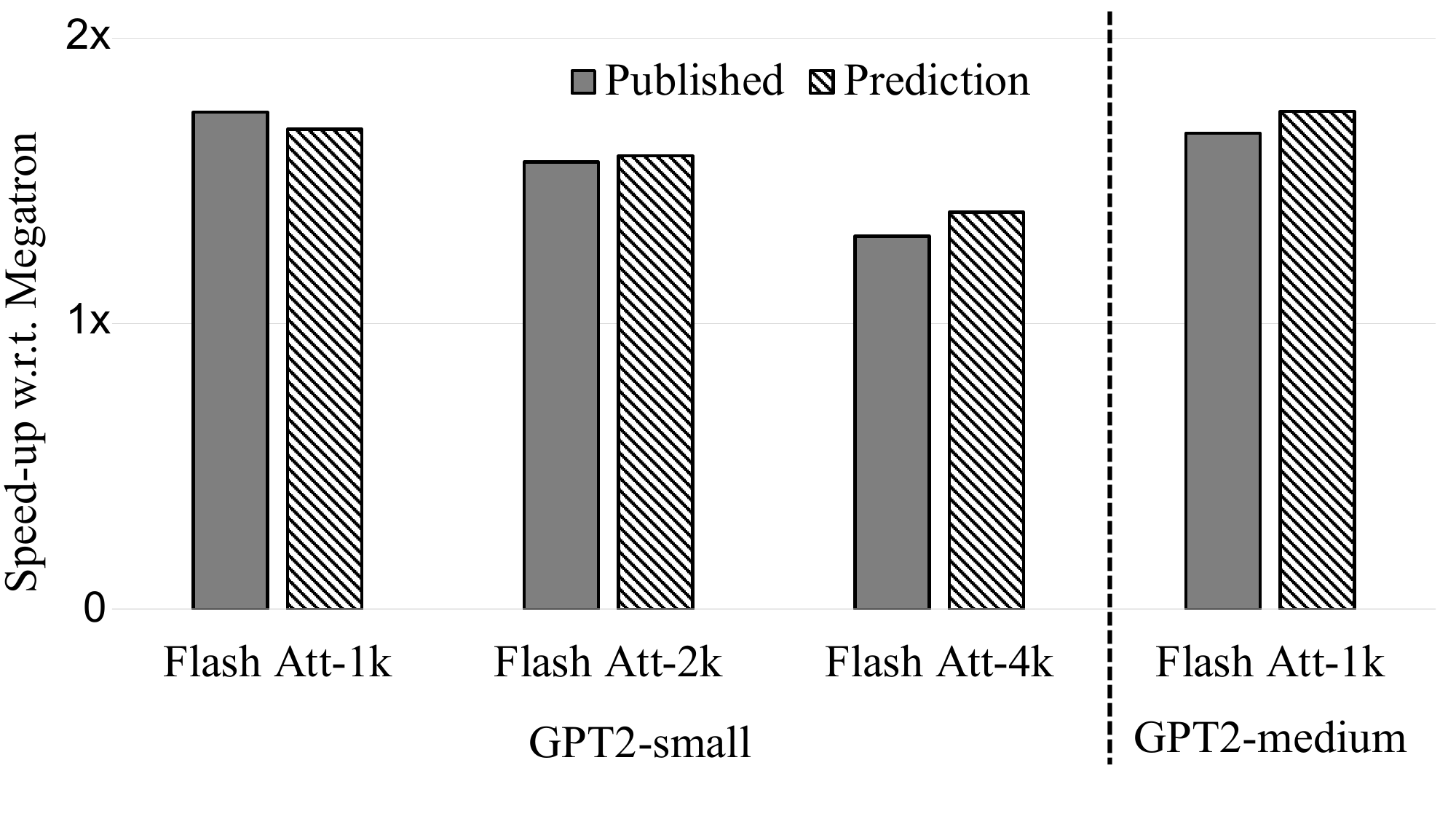}
    \caption{Validation of FlashAttention on 8xA100-40GB: the speed-up with FlashAttention technique for two different GPT2 models with growing context length from 1k-4k versus the Megatron implementation with a fixed context length of 1024.}
    \label{fig:fa}
\end{figure}
\section{Validation}

In this section, we validate our modeling methodology with published data in the literature. 
\begin{figure}
    \centering
    \includegraphics[width=\linewidth]{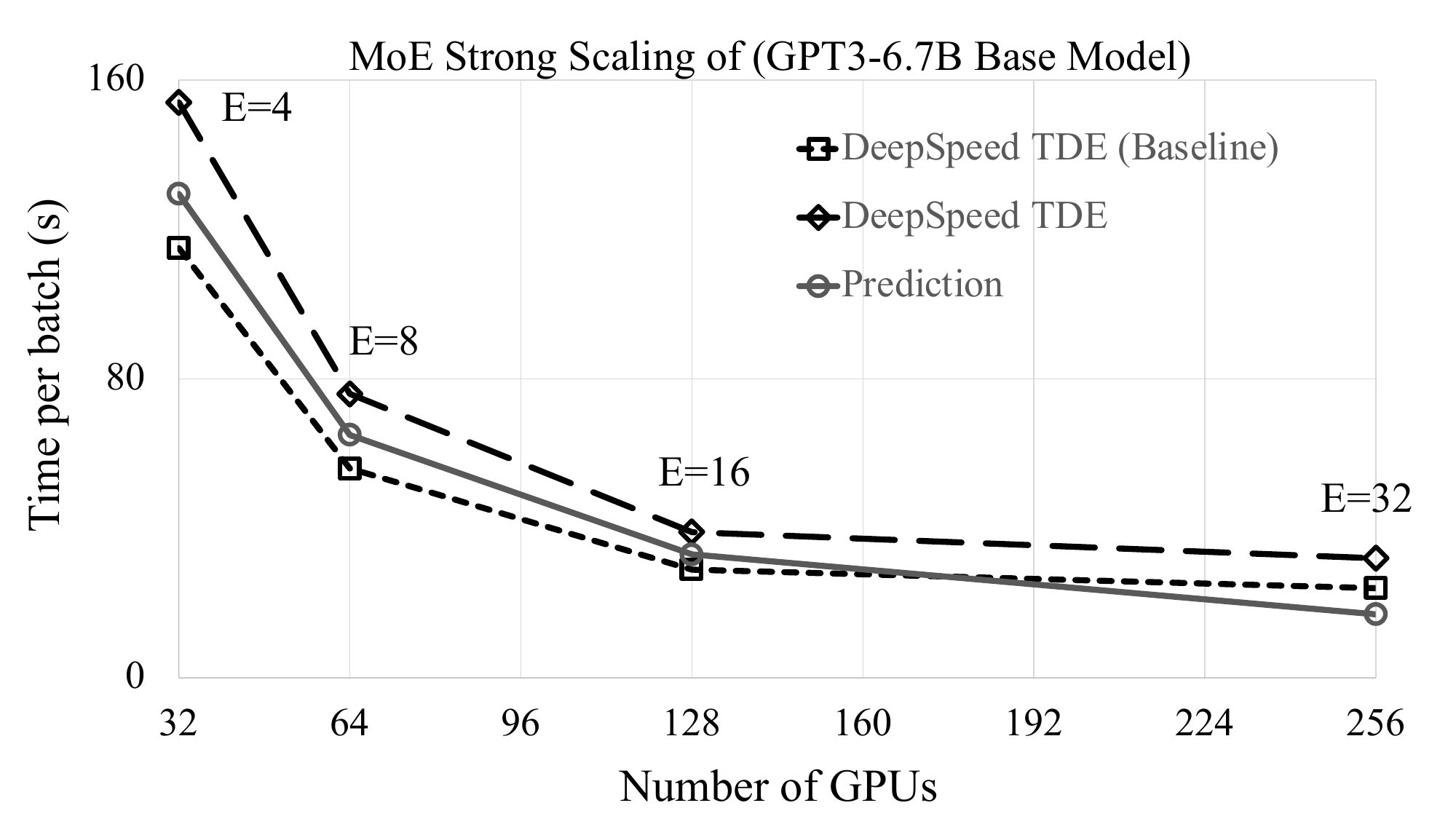}
    \caption{Validation of MoE: strong scaling of the GPT3-6.7B baseline model with varying number of experts and GPUs. Batch size is set to 1024. The solid line is our prediction and the dotted lines are obtained from \cite{DeepSpeedMoE}.}
    \label{fig:moe}
\end{figure}
\begin{figure}
    \centering
    \includegraphics[width=\linewidth]{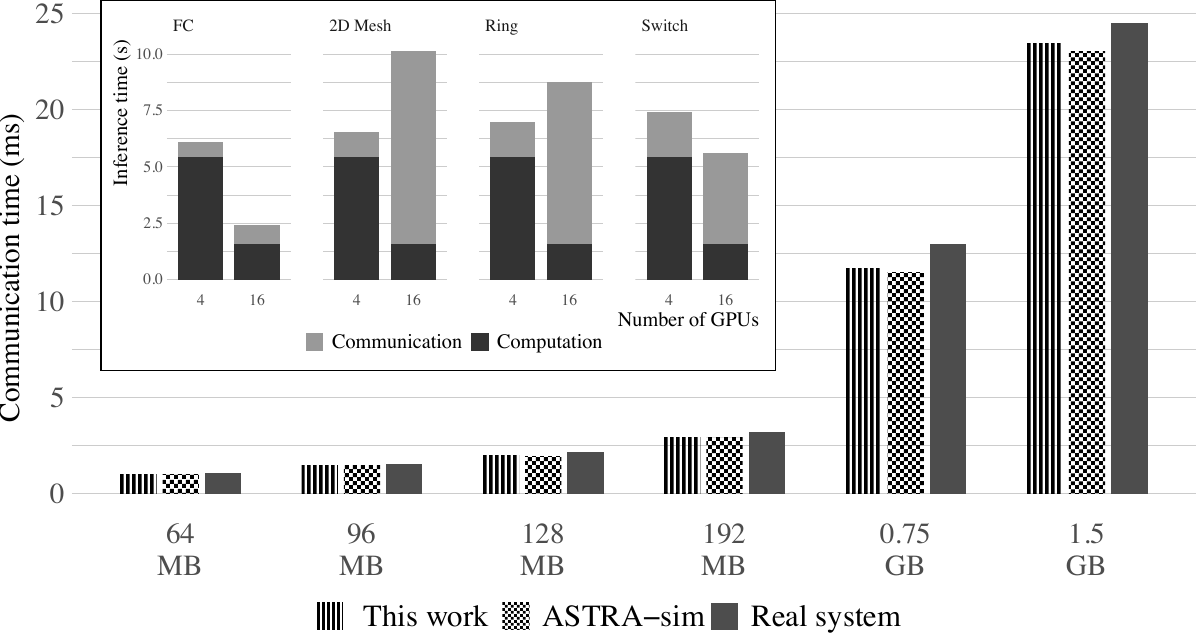}
    \caption{Validation of the network model. Inset showing the impact of topology on inference performance of Llama-70B model}.
    \label{fig:nw_valid}
\end{figure}

\subsection{FlashAttention}
To validate our model for FlashAttention-v1, we choose two different LLM models (GPT2-small with 125M parameters and GPT2-medium with 355M parameters) along with different context lengths from \cite{flashattention}. We configure the model for a system with 8 A100-40GB GPUs connected with NVLink gen-3 using data parallelism. The estimated speed-up is compared with the Megatron implementation of the corresponding GPT2 models whose context length is set to 1024. Fig.~\ref{fig:fa} illustrates the results. Depending on the setup, the speed-up varies from 1.3x-1.7x and the error margin of our prediction is within 8\%. The communication overhead due to DP is quite small in our setup, thus, the speed-up in compute time drives the effective speed-up. Note that the attention computation amounts to more than 60\% of the total time. Although FlashAttention drastically reduces the execution time of the attention computation by making the kernel compute bound, there are other compute chunks that remain unchanged: the KQV computation for all the heads, the projection step, and the MLP block. Thus, the upper bound of the speed-up is expected to be the order of 2x. It is important to note that, with FlashAttention, the same base model with 4k context length is 1.3-1.4x faster than the implementation without FlashAttention of 1k context length. Thus, we also validate the trend that for longer contexts, FlashAttention tends to be more beneficial ($>$ 3x speed-up for 4k context length).

\subsection{Mixture of Experts}
To validate the performance model for MoE, we reproduce the data published in DeepSpeed-TED with hybrid parallelisms \cite{Singh_2023}. We present the strong scaling of a 6.7B Base Model with B = 1024 as the number of experts per GPU is varied. It is worth noting that even though the model size scales with the increasing number of experts, the total number of floating operations in training does not change. The actual experiments were run on V100 nodes of the Summit supercomputer. Summit features 6 NVIDIA V100 GPUs per node, each with 16 GB of memory and a peak half-precision performance of 125 TFLOP/s. The GPUs support bidirectional communication with a peak intra-node bandwidth of 50 GB/s via NVLink and an inter-node bandwidth of 25 GB/s via InfiniBand. We model this system architecture within our framework.
\begin{figure}
    \centering
    \includegraphics[width=\linewidth]{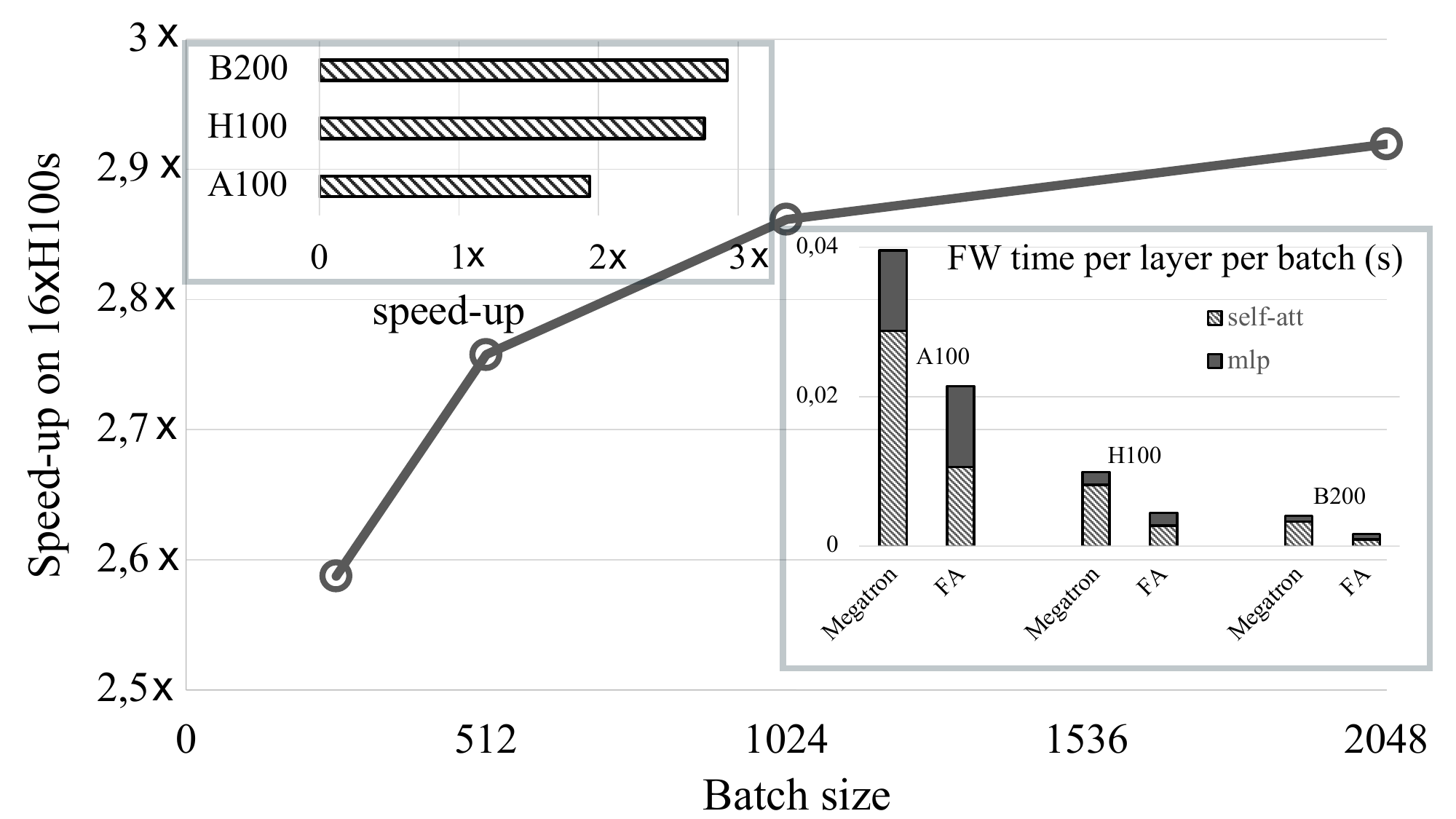}
    \caption{The speed-up due to FlashAttention as a function of batch size. Top-left panel: The speed-up for different GPUs for B = 2048. Bottom-right panel: the ratio between attention and mlp computation for different GPUs with and without FlashAttention.}
    \label{fig:fa_cs}
\end{figure}

The experiments in \cite{Singh_2023} covers two cases: I) the baseline hybrid TP-DP-EP parallelism, and II) with two optimizations: duplicate token dropping to reducing communication volume and communication-aware activation checkpointing. Our model does not account the impact of load-imbalance and the optimizations above. Hence, we expect our prediction to be inline with the baseline implementation. Our prediction lies between the two bounds of \cite{DeepSpeedMoE} as presented in Fig.~\ref{fig:moe} (time to execute a batch versus number of GPUs). For these experiments, we fix TP = 4 and vary the DP$_{\rm MHA}$ (= EP) degree. The performance tend to saturate since the communication overhead starts becoming considerable when more GPUs are involved. We see a slight departure from the trend when number of GPUs is 256. This is primarily due to the fact that we underestimate the impact of load imbalance when number of experts/GPUs becomes higher.

\subsection{Network modeling}
Validating the network model is performed through running an all-reduce collective of varying sizes between 64 MB and 1.5 GB on 16 NVIDIA V100 GPUs with 150 GB/s NVLink GPU-to-GPU bandwidth. This setup is the recreation of ASTRA-sim2.0 \cite{ispass-23-won} validation, for comparison purposes. The results in Fig. \ref{fig:nw_valid} show that the modeled communication times in this work lie between the results from ASTRA-sim and measurements of the real system. The difference between the results of this work and those of ASTRA-sim originates mainly from the introduction of switching latency. 



\begin{figure}
    \centering
    \includegraphics[width=\linewidth]{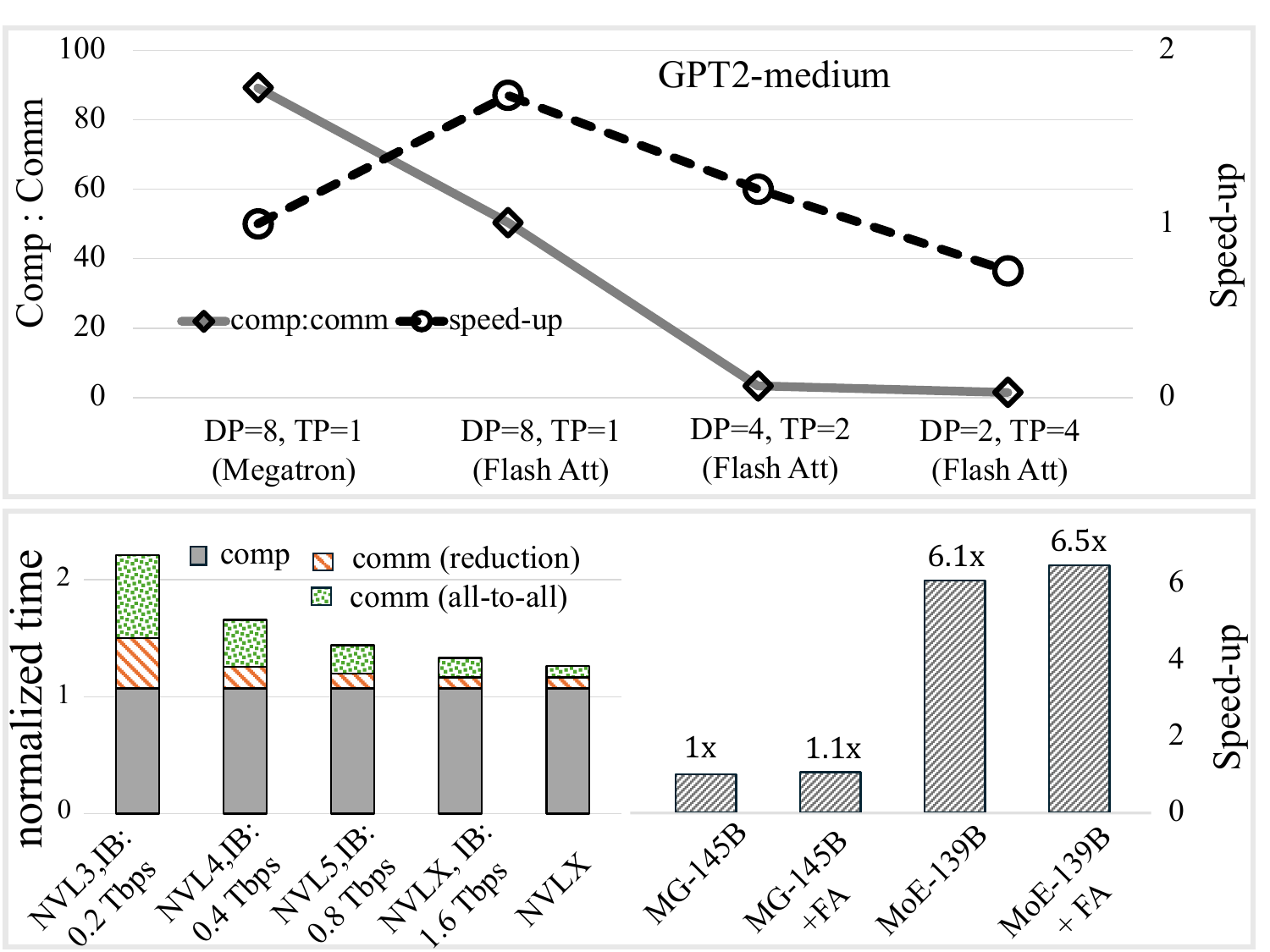}
    \caption{Top: impact of hybrid parallelization (DP + TP) on performance of FlashAttention and the correlation with compute to communication ratio. Bottom-left: impact of network bandwidth scaling on MoE performance. Bottom-right: speed-up with MoE (base GPT3-13B) when compared to an equivalent Megatron model (MG) and the impact of combining FlashAttention and expert parallelism. NVL: NVLink, IB: InfiniBand}
    \label{fig:fa_moe}
\end{figure}
\section{Case Studies}
\subsection{FlashAttention case study} 
In this section, we perform what-if analysis with FlashAttention using our performance model. We primarily explore ways to enhance the performance benefits of FlashAttention. Especially, we study how the speed-up with respect to the Megatron implementation scales as a function of batch-size \& GPU generation. In addition, we provide insights behind the speed-up and explore the impact of different parallelization choices. We consider GPT2-xl model with 1.5B parameters and a context length of 2048 on a system of 16 H100s connected via fourth generation NVLink. As shown in Fig.~\ref{fig:fa_cs}, the speed-up due to FlashAttention grows monotonically with increasing batch size (B). As B gets larger, the matrices in attention computation becomes skinnier or taller hurting the data reuse from HBM in the standard implementation. On the other hand, the FlashAttention implementation is agnostic of the above issue since it relies on SRAM tiling bypassing direct data movement to/from HBM. The top-left inset in Fig.~\ref{fig:fa_cs}, denotes the speed-up of the same model on different GPUs (A100, H100 and B200) for B = 2048. Note that we use the same precision for all the three GPU generations. Since FlashAttention turns the memory bound GEMM operations into compute bound ones, the attention computation benefits from the higher flop rate of newer GPU generations. HBM bandwidth also gets better while going from A100 to H100 to B200 -- but the flop rate growth is faster than that of the HBM bandwidth benefiting FlashAttention more. To rationalize the speed-up seen for FlashAttention accross different GPU generations, we look closely what fraction of time is spent in attention computation when compared to the MLP block for both the Megatron and FlashAttention implementation. In all cases, the attention computation (including the QKV computation) takes more time than MLP for the Megatron implementation due to the HBM bandwidth boundedness of most of the operations. Upon using FlashAttention, the time spent in attention computation reduces significantly but the time for MLP block remains unchanged -- see the bottom-right inset of Fig.~\ref{fig:fa_cs}. 

\textbf{\underline{Key takeaway}--} \textit{FlashAttention is more beneficial when memory bandwidth becomes a larger bottleneck in the standard implementation, for example, while using larger batch sizes (less reuse of HBM accesses) or newer GPU generations (the increase in peak flop rate is typically larger than the increase in memory bandwidth).
}

    Next we analyze the impact of hybrid parallelization choices on the speed-up. In Fig.~\ref{fig:fa_moe} top panel, we show the impact of varying DP and TP degree on FlashAttention's speed-up compared to the Megatron case with DP = 8 and TP = 1. The shown estimations are for GPT2-medium on an 8xA100-40GB system as described before. As expected the FlashAttention implementation with DP=8 and TP = 1 leads to a $\sim$ 1.7x speed-up. However, FlashAttention starts to show diminishing benefits as we increase TP from 1 to 2 to 4. The reason is the growing communication cost primarily, due to the all-reduce collectives of TP. As the ratio between compute and communication goes down, the overall speedup due to FlashAttention also decays since it accelerates only a specific part of the compute. However, a potential solution to mitigate the communication overhead is to employ fully-sharded data parallelism (FSDP) instead of TP. FSDP effectively overlaps communication with computations, thereby preserving the performance benefits of FlashAttention.

\textbf{\underline{Key takeaway}--} \textit{
FlashAttention's benefits are reduced, when the communication overhead is no longer negligible (at scale). Parallelism strategies, such as FSDP, can mitigate the overhead and preserve the benefits.}

     \subsection{Case study with MoE and FlashAttention} 
     Here, we study the impact of bandwidth scaling on MoE's performance, implications of having FlashAttention along with pipeline parallelism, and further, mixing FlashAttention with expert parallelism for an equivalent MoE model. 

     Bottom-left panel of Fig.~\ref{fig:fa_moe} we show the impact of network bandwidth scaling (towards optical interconnect) on reduction and all-to-all communication for a MoE model with 139B parameters (base model of GPT3-13B) on 256 GPUs. We set B=512, DP=EP=32 and TP=8, keeping the compute fixed to A100-80GB. The relative fractions are in agreement with \cite{DeepSpeedMoE} and the data are normalized by the fixed compute time. NVL-i and IB stand for NVLinlk gen-i for intra-node and InfiniBand for inter-node network. NVL-X is a hypothetical next gen NVLink with a bandwidth = 2x NVLink gen-5. As we can see that both communications goes down as the network bandwidth scales. When both intra and inter node bandwidths are set at the same value, reduction and all-to-all communication becomes almost equal and the gain due to higher bandwidth tend to saturate -- especially, the massive inter-node bandwidth shows a diminishing return. Since the communication time becomes negligible, we see improved speed-up ($\sim$ 1.2x) with FlashAttention at this limit. 

\textbf{\underline{Key takeaway}--} \textit{
The class of MoE models is an effective way to increase model size and accuracy with minimal or no additional compute cost. However, the system-wide all-to-all communication overhead can be very high $>$ 30\%. Scaling inter-node  network bandwidth helps reducing the communication overhead.}

     Next, we study the impact of including FlashAttention within MoE (bottom-right panel of Fig.~\ref{fig:fa_moe}). For the base case, we choose Megatron-145B model training with 300B tokens on 32 nodes, each with 8 accelerators modeled after the A100-80GB GPU, with an intra-node network using third generation NVLink. The inter-node network is modeled after InfiniBand with 200Gb/s. For this case, we set a batch size of 192 with TP = 8, DP = 8, and PP = 4 with  8 micro-batches. We make sure that there is no memory overflow and the predicted performance is 149 TFLOP/s (similar to what is reported in the Megatron paper\cite{eff_megatron}). As discussed in earlier section, the performance improvement due to FlashAttention in this setup with TP is expected to be minimal-- a modest 1.1x speed-up is presented in Fig.~\ref{fig:fa_moe} bottom panel. Next we use MoE-139B parameter model with 32 experts (each token is directed to only 1 expert). The underlying base model is GPT3-13B. For this experiment, we set a batch size of 512 with DP = EP = 32, TP = 8, PP = 1 (\# of GPUs = 256). The total parameter count of the MoE model is equivalent to the chosen Megatron model, however, it actively applies only 13B active parameters resulting in much less computational complexity, while keeping accuracy at the same level (following the scaling law\cite{kaplan2020scalinglawsneurallanguage}). Of course, there are overheads of a larger memory footprint and all-to-all communications due to MoE. In terms of performance, the MoE model achieves $\sim $ 6.1x speed-up compared to the Megatron model. Regarding the performance impact of implementing FlashAttention in the MHA block of the MoE model, we observe a further speed of $\sim$ 1.1x, which amounts to a $\sim$ 6.5x overall speed-up compared to the Megatron case without FlashAttention.  

\begin{figure*}[ht!]
\centering
    \begin{subfigure}{0.31\linewidth}
        \centering
        \includegraphics[width=\linewidth]{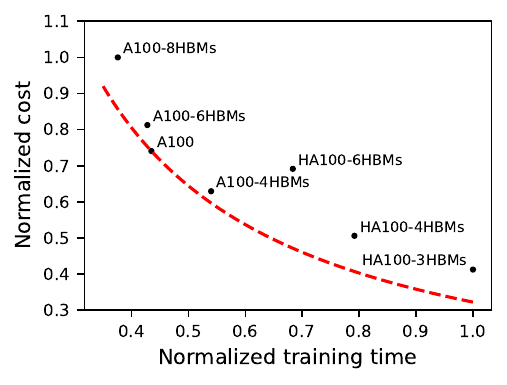}
        \caption{}
        \label{fig:pc_training}
    \end{subfigure}
    \begin{subfigure}{0.31\linewidth}
        \centering
        \includegraphics[width=\linewidth]{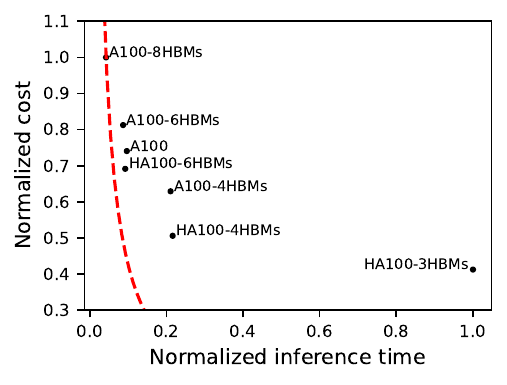}
        \caption{}
        \label{fig:pc_inf}
    \end{subfigure}
    \begin{subfigure}{0.31\linewidth}
        \centering
        \includegraphics[width=\linewidth]{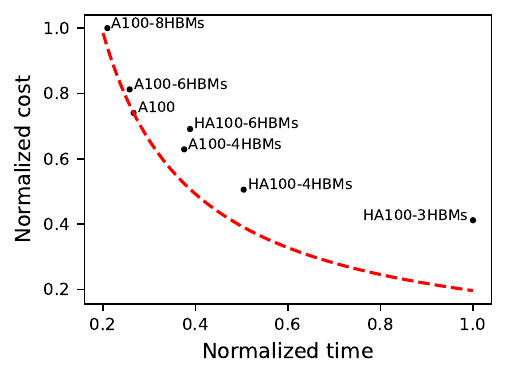}
        \caption{}
        \label{fig:pc_total}
    \end{subfigure}    
    \caption{Performance-cost tradeoffs for various designs based on Nvidia A100. The GPT3-175B model on 1024 GPUs is used in all the cases. The dashed line in red is an iso-product curve. (a) Cost vs training time; measured for one epoch. (b) Cost vs inference time; for serving 1B requests. (c) Cost vs combined training and inference time. The combined time is calculated as weighted normalized training time plus weighted normalized inference time. The weight factor is chosen as 0.5.}
    \label{fig:perf_cost}
\end{figure*}
\subsection{Case study on network topologies}
In this experiment, we created a system of 8 and 16 NVIDIA A100 GPUs connected by a variety of topologies. The total off-GPU bandwidth is divided by the number of links to determine the link bandwidth in each topology. The inset of Fig. \ref{fig:nw_valid} shows the communication and computation breakdown of running Llama 80B inference on this setup. This workload is latency sensitive due to the communication volumes involved. The fully-connected topology performs the best, followed by the switch, as regardless of the number of GPUs, the mean number of hops in these topologies are always 1 and 2, respectively. Ring and 2D mesh do not perform as well, especially at larger system size. This is mainly because the number of hops to complete a collective gets larger with the system scale. 2D mesh is impacted worse-off than the ring because the link bandwidth is lower in the 2D mesh compared to the ring.

\subsection{Performance-cost case study} 
Generally, scaling up a system leads to better performance. However, it could incur significant cost. The achieved performance gain might not compensate for the rapid rise in the cost. The integration of the cost model allows us to evaluate the performance-cost trade-off. We show a case study, inspired by the NVIDIA A100 GPU design, to demonstrate the performance-cost trade-off when selecting a number of HBM stacks for an accelerator design. 

The A100 GPU consists of a full reticle-sized logic die and five HBM stacks. By scaling either the logic or memory, we come up with various configurations for comparisons. Firstly, the size of the logic die is kept unchanged, but the number of HBM stacks is varied from 4 to 8. This results in three different configurations denoted as A100-4HBMs, A100-6HBMs, and A100-8HBMs, respectively. Secondly, the size of the logic die is halved and the number of HBM stacks is varied from 3 to 6, resulting in the following configurations: HA100-3HBMs, HA100-4HBMs, and HA100-6HBMs. We estimate the cost and performance for the GPT3-175B model for training and inference on a 1024-GPU cluster, where each node has eight GPUs, following a node architecture inspired by the DGX A100 super-pod reference design \cite{amperedgx}. Table \ref{tab:perf-cost} lists the batch size and parallelism parameters. The backend parameters (e.g., cost per $mm^2$, defect density, lithography percentage and NRE cost) in the cost model for A100 GPUs at 7nm process node are obtained from IMEC research fabs \cite{24-ale}. The die size of A100 GPU and HBM2E are extracted from TechInsights reports \cite{techinsights}. 

Fig. \ref{fig:perf_cost} shows the performance and cost results for all the configurations. In Fig. \ref{fig:pc_training}, training time is measured for one epoch. In Fig. \ref{fig:pc_inf}, inference is performed in one node, so only tensor parallelism is used, however, we show the time required to serve one billion user requests, using the whole cluster. An iso-product curve is plotted as the red dashed line. For training, increasing the number of HBM stacks leads to performance gain, but not as significant as the cost rise, because training is mainly compute-bound. However, decreasing the number of HBM stacks can also hurt the performance more than its cost savings. On the other hand, by halving the size of the logic die, the performance degradation is more substantial than the cost reduction, because HBM stacks are generally costly. The optimal configuration is the one used in the A100 GPU. Since inference is memory-bound, increasing the number of HBM stacks leads to more significant performance gain. Reducing the size of logic die can effectively lower the cost with little performance loss. In order to select the best design for both training and inference, we normalize the training and inference time separately and equally weigh them to derive a normalized time as shown in Fig. \ref{fig:pc_total}. Based on this performance metric, the A100 GPU has an optimal design, which is in line with industry trends. 

\textbf{\underline{Key takeaway}--} \textit{
Hypothetically halving the die area of an A100 GPU, while keeping the same amount of HBM stacks, is an effective way to reduce cost, while maintaining inference performance. However, during training, it reduces performance more than it saves on cost.}

\textbf{\underline{Key takeaway}--} \textit{Increasing HBM stacks to boost memory bandwidth of an A100-like GPU is expensive. This is still cost-effective to enhance performance for inference but not for training. Choosing the right amount depends on the intended use of the accelerator cluster, with equal usage for training vs.inference favoring the design choice made in the A100 GPU.}

\begin{table}[]
\centering
\caption{Experimental settings for performance-cost case study}
\label{tab:perf-cost}
\begin{tabular}{|l|l|lllll|}
\hline
\multirow{2}{*}{} & \multirow{2}{*}{Batch size} & \multicolumn{5}{c|}{Parallelism}                                                                            \\ \cline{3-7} 
                  &                             & \multicolumn{1}{l|}{DP} & \multicolumn{1}{l|}{TP} & \multicolumn{1}{l|}{PP} & \multicolumn{1}{l|}{SP} & EP  \\ \hline
Training          & 256                         & \multicolumn{1}{l|}{16} & \multicolumn{1}{l|}{8}  & \multicolumn{1}{l|}{8}  & \multicolumn{1}{l|}{8}  & n/a \\ \hline
Inference         & 32                          & \multicolumn{1}{l|}{1}  & \multicolumn{1}{l|}{8}  & \multicolumn{1}{l|}{1}  & \multicolumn{1}{l|}{8}  & n/a \\ \hline
\end{tabular}
\end{table}

\section{Summary}
The primary advantage of an analytical performance model is its ability to facilitate early design exploration that would be too costly to implement and test on actual hardware. This paper presents a detailed study of performance modeling for state-of-the-art LLM implementations at scale, building on top of an existing analytical model. We validate our predictions using published results obtained on real hardware. Using the analytical model, we examine how the performance of FlashAttention varies with factors such as batch size, context length, 5D parallelization strategies, and GPU generations. We also explore the impact of bandwidth scaling on the communication overhead of MoE models and investigate the performance benefits of combining FlashAttention with expert parallelism. Our analysis focuses on FlashAttention-v1 and DeepSpeed's TED baseline implementation. The cost model we employ is limited to chiplets and does not account for full-scale data-center infrastructure. Future work will expand on these aspects. We believe this framework lays the foundation for a comprehensive performance and total cost of ownership analysis platform for large scale AI systems.



\end{document}